\documentclass[12pt,preprint]{aastex}

\def \nuc#1#2{\relax\ifmmode{}^{#1}{\protect\text{#2}}\else${}^{#1}$#2\fi}

\begin{document}

\title{A Model For the Formation of High Density Clumps in Proto-Planetary Nebulae}

\shorttitle{High Density Clumps in PPN}

\author{Patrick A. Young\altaffilmark{1},
J. L. Highberger\altaffilmark{1,2}, David Arnett\altaffilmark{1}, \&
L. M. Ziurys\altaffilmark{1,2}}

\altaffiltext{1}{Steward Observatory, University of Arizona, 933
N. Cherry Avenue, Tucson AZ 85721}

\altaffiltext{2}{Department of Chemistry, University of Arizona, 933
N. Cherry Avenue, Tucson AZ 85721}

\email{payoung@as.arizona.edu, jaimeh@as.arizona.edu,
darnett@as.arizona.edu, lziurys@as.arizona.edu}

\begin{abstract}
The detection of NaCl at large radii in the Egg Nebula, CRL 2688,
requires densities of $10^7$-- $10^8\ {\rm cm^{-3}}$ in a thick shell
of $ r \sim {\rm a\ few} \times 10^{17}\ {\rm cm}$. To explain these
results, a mechanism is needed for producing high densities at a
considerable distance from the central star. In two dimensional
simulations of the interaction of the fast wind with an ambient
medium, the material becomes thermally unstable. The resulting clumps
can achieve the requisite conditions for NaCl excitation. We present
2D models with simple physics as proof-of-principle calculations to
show that the clumping behavior is robust. Clumping is a natural
outcome of cooling in the colliding wind model and comparable to that
inferred from observations.

\end{abstract}
\keywords{astrochemistry --- hydrodynamics --- stars: AGB and post-AGB --- stars: winds, outflows}


\section{OBSERVATIONAL MOTIVATION}

Mass loss from Asymptotic Giant Branch (AGB) stars is a major avenue
by which heavy elements enter the interstellar medium
\citep{kwok00}. In addition to the initial phase of mass loss on the
AGB, a second episode is thought to occur in the post-AGB stage which
is characterized by high velocity winds that collide with the remnant
envelope. This wind-wind interaction is thought to generate drastic
departures from spherical symmetry in the remnant shell
(i.e. bi-polar), as well as very hot gas, as traced by
vibrationally-excited lines of ${\rm H_2}$\ \citep{sah98,cox00,her02}.

Recently, \citet{hig03} detected NaCl and NaCN in the post-AGB star
CRL 2688 in an optically thin extended shell of radius $\sim
10$--$12\arcsec$\ (a few $\times 10^{17}\ {\rm cm}$ at $\sim 1$\
kpc). This result is surprising, as it corresponds to a much larger
physical radius than is observed in the AGB star IRC+10216
\citep{gln97}. The NaCl lines in CRL 2688 do not show the
high-velocity tails characteristic of post-AGB mass ejection.
Instead, the cusp-shaped line profiles of NaCl, observed with the IRAM
30m telescope, indicate an extended, shell-like distribution for NaCl
coincident with the low velocity AGB remnant wind. 
A similar interpretation holds for NaCN.  The observed transitions of
NaCl and NaCN in CRL 2688 require densities of $n({\rm H_2}) \sim
5\times 10^6\ {\rm cm^{-3}}$ to $n({\rm H_2}) \sim 10^8\ {\rm
cm^{-3}}$ for $T_K \sim 50\ {\rm K}$ to be excited. These values are
inconsistent with an undisturbed AGB remnant wind \citep{hig03}.

We propose that clumping of material near the interface between the
slow AGB wind and the fast post-AGB wind can reach the requisite
densities for excitation of the observed transitions of NaCl and
NaCN. This clumping is driven by a thermal instability arising in a
hydrodynamic shock. In this letter we present two dimensional
proof-of-principle calculations which show that the clumping process
occurs and is robust for conservative estimates of proto-planetary
nebula (PPN) conditions, and that the clumping can produce the
requisite densities.

\section{THEORY OF CLUMPING}

Modeling of the interaction of winds in post-AGB stars and PPN by using
one-dimensional spherically symmetric shocks precludes
inclusion of physics relevant to the morphology and conditions in the
interaction region. In particular, the assumption of spherical
symmetry prevents material from fragmenting into small clumps which
can have significantly enhanced density and a more radially extended
distribution than a simple shock.

The inclusion of radiative cooling in the gas physics gives the
possibility of a thermal instability. In this scenario, the clumps are
pressure-confined. Cooling is usually non-linear with density (i.e.,
following \citet{ost}, $\Lambda \varpropto \rho^2$ for free-free
emission or collisionally excited line radiation). Small density
anisotropies amplified by a shock or shock precursor cool much more
efficiently than the surrounding medium and are no longer in pressure
equilibrium.

The size of the clumps is set approximately where the sound travel
timescale is commensurate with the cooling timescale. Density
variations across larger scales will not have time to be smoothed out
before the cooling-driven instability sets in. 

\section{MODELS OF CLUMPING}

\subsection{Physical Assumptions}

The simulations were produced with a version of the PROMETHEUS PPM
hydrodynamics code optimized for stellar wind/CSM interactions
\citep{mfa91, ma95}. All calculations were performed in two dimensions
on a spherical grid. A $300\times 140$ grid was used for testing
different cooling curves and CSM and wind conditions. The calculations
for the most reasonable combination of parameters were repeated with
$529\times 426$ and $1000\times 806$ grids.

The equation of state is the ideal gas law with ionization from
electron collisions and the ambient radiation field and with
recombination.  Two sets of cooling curves were used. Both sets
include free-free emission from \citet{ost} and high temperature
cooling from \citet{kaf73}. The second set of curves has additional
cooling for collisionally excited line radiation and fine structure
recombination lines \citep{ost} and rotational transitions of CO
\citep{hm79}. The collisionally excited line radiation and fine
structure line curves are simple analytic functions which were
designed to have the qualitatively correct functional form, and are
normalized to the peak values of the curves in \citet{ost}. No attempt
was made to reproduce the detailed physics of the cooling, merely to
reproduce the correct order of magnitude values. The results are
largely insensitive to the form of the cooling curve, so long as
cooling is present, so we did not attempt to improve our curves. In
some circumstances there is a numerical instability in PPM codes
related to cooling \citep{sut03}, however examination of our resolved
simulations indicates we are not in this regime.

Since these simulations were motivated by the detection of NaCl in CRL
2688, we will briefly review the physical conditions in this
object. Conditions in the circumstellar envelope have been determined
from rotational line studies of CO. The inner ``superwind''
corresponds to a mass loss rate of $\dot{M} \sim 10^{-3}\ M_{\sun}\
{\rm yr^{-1}} $ \citep{you92} and a velocity of $v \sim 60-200\ {\rm
km\ s^{-1}}$ \citep{her02}. The surrounding circumstellar material
(CSM), produced by AGB mass loss, has $n({\rm H_2}) \sim 5\times 10^5\
{\rm cm^{-3}}$, $T \sim 50\ {\rm K}$, and $v \sim 20\ {\rm km\
s^{-1}}$ \citep{her02}. Shock zones form in the region where the fast
wind interacts with the surrounding material. Emission from CO and
${\rm H_{2}}$ has been seen in this shocked region, located $\sim
6$--$7\arcsec$ from the central star and distributed in a clover-leaf
shaped distribution \citep{sah98, cox97, cox00}.  In the optical and
IR CRL 2688 shows a distinct bipolar morphology like many other PPN
and PN \citep[e.g.][]{sah98}.

We chose to use a somewhat more conservative (with regard to the
clumping instability) set of conditions for the CSM and fast wind. For
the CSM we used $n({\rm H_2}) \sim 1\times 10^5\ {\rm cm^{-3}}$ and
$T_K \sim 50\ {\rm K}$. The CSM was given an enhanced equatorial
density that increases with polar angle as $sin^{64}\theta$ and a
maximum enhancement in the density of 50\% \citep{ma95, nrc95}. (This
geometry was chosen since a large fraction of PN and CRL 2688 in
particular are bipolar in shape, but the results turn out to be
relatively insensitive to asymmetries at this level
.) For
the inner fast wind we used a mass loss history produced by the
stellar evolution code TYCHO \citep{ymal01, ykra03}. The median values
produced by this mass loss history were $\dot{M} \sim 2\times 10^{-4}\
M_{\sun}\ {\rm yr^{-1}} $ and a differential velocity between the
components of $v \sim 30\ {\rm km\ s^{-1}}$. Because of the short
timescales in question, these values were approximately constant over
the duration of the simulation. Higher densities (or mass loss rates)
in either component tend to promote clumping. The degree of clumping
was relatively insensitive to the difference in velocity of the
components up to $v \sim 80\ {\rm km\ s^{-1}}$, though the radius of
onset of clumping was larger since the expansion timescale was reduced
relative to the cooling timescale. Above $80\ {\rm km\ s^{-1}}$ the
temperatures in the shock exceeded $10^4$ K, which is higher than the
observed vibrational temperature of the shocked gas in CRL 2688. (From
their measurements of ${\rm H_2}$, \citet{cox97} find $T_{vib} \sim
3000\ {\rm K}$.) The inner boundary of the simulation was set at
$5\times 10^{16}\ {\rm cm}$.

A few further caveats must be kept in mind. First, the simulations are
two dimensional. This means that collapse into clumps is constrained
to two dimensions; in the third the material forms annuli around the
axis of symmetry. In a real system the material can collapse in three
dimensions, resulting in larger density enhancements, more efficient
fingering (since material is being displaced along a small-cross
section plume rather than an entire annulus), and a more complex
morphology. (Fingering here refers to the common behavior of
finger-like protrusions developing at an unstable interface.) Secondly,
the wind and CSM are plasmas, so they can and probably do support
magnetic fields, the effects of which are considerably more difficult
to predict {\it a priori}. Third, the morphology of a real
proto-planetary nebula is considerably more complex than the smoothly
varying distribution examined here. The presence of molecular
outflows, jets, multiple AGB wind components and large scale density
perturbations will change the distribution and morphology of
clumps. In fact, multiple molecular outflows have been observed in CRL
2688 \citep{cox00}.

\subsection{Results}

We find that for the conditions described in section 3.1, the highest
density in the simulation increases at the beginning of the simulation
as the shock establishes itself, then decreases for a time before
reaching a minimum as cooling takes over from spherical divergence as
the primary process controlling the density evolution. The density
thereafter increases in the clumps until the shock front moves off the
grid. 

The maximum densities and the radii at which they are achieved
emphasize the importance of grid resolution to this simulation. As
discussed in \S 2, the physical size of the clumps should be quite
small. The size of the clumps, and correspondingly their maximum
density, is limited by the grid in the low and medium resolution
simulations. The ratio of maximum densities achieved in two
underresolved simulations scale as the square of the increase in
radial resolution between the simulations. The onset of clumping
occurs much earlier for higher resolution, as well. This scaling no
longer holds between the medium and high resolution simulations. The
maximum densities and radius of onset of clumping do not differ by
more than 10\% at any point, indicating that the medium resolution
simulation is slightly under-resolved or barely resolved and the high
resolution over-resolved.

In the medium resolution simulations, a maximum density of $n =
3.4\times 10^6 \ {\rm cm^{-3}}$ occurs at $r \sim 1.5\times 10^{17}\
{\rm cm}$, and clumping is already developing. The density drops to $n
= 2.8\times 10^6\ {\rm cm^{-3}}$ at $r \sim 2.5\times 10^{17}\ {\rm
  cm}$ and returns to near its peak value by $r \sim 3.5\times
10^{17}\ {\rm cm}$ at the end of the simulation. In the high
resolution simulation, a local maximum density of $n = 3.7\times 10^6
\ {\rm cm^{-3}}$\ is reached at $r \sim 1.5\times 10^{17}\ {\rm
  cm}$. The following minimum of $n = 3.0\times 10^6 \ {\rm cm^{-3}}$\
occurs at $r \sim 2.6\times 10^{17}\ {\rm cm}$\ and density rises to
$n = 3.9\times 10^6 \ {\rm cm^{-3}}$\ by the end of the simulation, at
which point it is still increasing. 

In both cases the timescale for the full development of clumping is a
few hundred years, which is approximately the timescale for the
interaction of the fast wind with the CSM in CRL 2688 according to
${\rm H_2}$ observations \citep{sah98}. If we define clumps as a
plateau of locally maximum density in the simulation, the physical
sizes at $r \sim 1.5\times 10^{17}\ {\rm cm}$\ range around $\geq
0.5\times 10^{15}\ {\rm cm}$, with masses of approximately $\geq
0.5\times 10^{-5}\ M_{\sun}$. (The resolution of the grid at this
radius is $1.8\times 10^{15}\ {\rm cm}$.) The density falls off to
near the ambient over a similar distance. The clumps also have long
``tails'' at lower density due to the fingering of the
instability. Figure 1 shows the fully developed clumping in the high
resolution simulation. The white line shows the position of the shock.

In order to ensure the size of the clumps is not a numerical artifact,
we ran simulations with 1\% random gaussian density fluctuations or a
sinusoidal 10\% density perturbation with an angular frequency of
$20/2\pi\ {\rm sr^{-1}}$. These perturbations should overwhelm
roundoff error at cell boundaries as seeds for the instability and
start fingering on much larger scales if the natural physical scale of
the clumps is larger than the grid size. No difference is seen in
clump sizes with and without the perturbations. To confirm that the
clumping is driven by a thermal instability, we performed a control
simulation with no cooling. Figure 2 compares simulations with and
without cooling. Without cooling no instability develops, even at
large radii. (The vortex at the bottom of the no cooling case is a
result of an interaction between the boundary conditions and the
equatorial density enhancement and does not appear in spherically
symmetric simulations.)  Both panels are on the same density
scale. The inner radius of the onset of clumping is a more difficult
problem. It can potentially be varied widely simply by assuming
different geometries for the circumstellar medium and conditions in
the CSM and fast wind at the beginning of the PPN phase.  The clumps
themselves, once formed, have a lifetime long compared to the
simulations, and may well move out to significantly larger
distances. At larger radii, even if the clumps are present, their
covering factor will be sufficiently small that beam dilution from a
single dish will make them once again unobservable. We must be careful
in interpreting the extent of the clumps from observations. Their true
distribution may be larger. We have demonstrated that the clumping can
happen, but other physics besides that of clump formation may
determine the {\it observed} scale.

The densities reached in the high resolution simulations exceed the
densities needed for excitation of the ${\rm J} = 7 \rightarrow 6$
transition of NaCl ($n = 3.4\times 10^6\ {\rm cm^{-3}}$ at 50 K), but
are a factor of several short of what is needed for higher
transitions. This is not problematic, and perhaps even desirable, for
two main reasons. First, the conditions in the CSM and fast wind are
conservative. Higher mass loss rates translate to increased
clumping. Second, the simulation is only in 2D, which only allows
fragmentation and compression of the clumps along two axes. Were these
annuli allowed to fragment in the third dimension the densities would
be enhanced further.

The structure of the shock is also worthy of note. The main shock
(characterized by a change in the sign of ${\bf \bigtriangledown \cdot
V }$ from expansion to compression and a temperature of $T \sim {\rm
a\ few} \times 10^3\ {\rm K}$) is interior to the clumping. The main
shock is preceded by a hydrodynamic precursor. The C abundance of the
fast wind was set to a factor of $10^3$ lower than that of the CSM to
provide a tracer of the Lagrangian motion of the material. Figure 3
shows the C abundance for the low resolution simulation at the same
timestep as the density plot. The C poor material has clearly moved
ahead of the main shock and shows a fingering pattern identical to
that seen in the density. Other mechanisms for the precursor are ruled
out independently. The code does not contain the physics for a
magnetic or cosmic ray precursor, and the temperatures in the shock
are too low to produce a photoionization precursor. It is interesting
to note that the measured radii of the NaCl and NaCN emission and the
shock front as measured by $\rm H_2$ are $\sim 10$--$12\arcsec$ and
$\sim 6$--$7\arcsec$, respectively \citep{hig03, sah98, cox97}. This
difference seems to indicate that the high densities occur {\it
outside} the strong shock, as we see in these simulations. 

\section{DISCUSSION}

New molecular observations of NaCl and NaCN in the proto-planetary
nebula CRL 2688 imply extreme densities in a shell at large radii from
the central star. We investigate the behavior of a thermally unstable
shock produced by a fast superwind impacting CSM from previous stages
of mass loss with a two dimensional hydrodynamics code. When even
simple cooling is included, the hydrodynamics initiates a thermal
instability in a shock precursor which leads to the formation of high
density clumps. These clumps can achieve the requisite densities at
large radii from the star. The clumping mechanism is robust, occurring
even for conservative estimates of PPN conditions.

Observations in the optical/IR may offer insight into
this model. The clump sizes we see in simulations are similar to those
observed for cometary globules in the Helix Nebula \citep{oh96}. This
does not necessarily imply that the clumps which we posit for PPN are
the precursors of PN clumps. However, the composition and the cooling
curves should be nearly the same for both cases. If the cometary
globules are the product of a thermal/hydrodynamic instability, which
is plausible but not certain \citep{hug02}, then their characteristic
size should be similar to what we predict for clumps in our
simulations. 


Ideally we would like to identify the structures in CRL 2688
associated with the molecular emission. The most likely scenario is
that the sodium molecules are associated with material traced by
vibrationally-excited ${\rm H_2}$, which exhibits a clumpy,
clover-leaf-shaped distribution \citep{cox00}. The emission suggests
that this is shocked gas from a wind-wind interaction. It is from this
region that we take our initial conditions. The inferred shock
temperatures are consistent with our model, and the geometries can be
accommodated by assuming a more structured CSM and fast wind. Our
models indicate that the clumping behavior is robust for the physical
conditions associated with the shock. The simulations indicate that
the Na molecule emission should be in a precursor to the shock. Just
outside the ${\rm H_2}$\ is the most likely location for the NaCl and
NaCN emission if the clumping mechanism is indeed responsible for
creating the requisite conditions for excitation. This agrees very
well with the observational evidence available for the location of the
Na molecules. The measured radii of the NaCl and NaCN emission and the
$\rm H_2$ are $\sim 10$--$12\arcsec$ and $\sim 6$--$7\arcsec$,
respectively \citep{hig03, sah98, cox97}

\clearpage


\begin{figure}
\figurenum{1}
\plotone{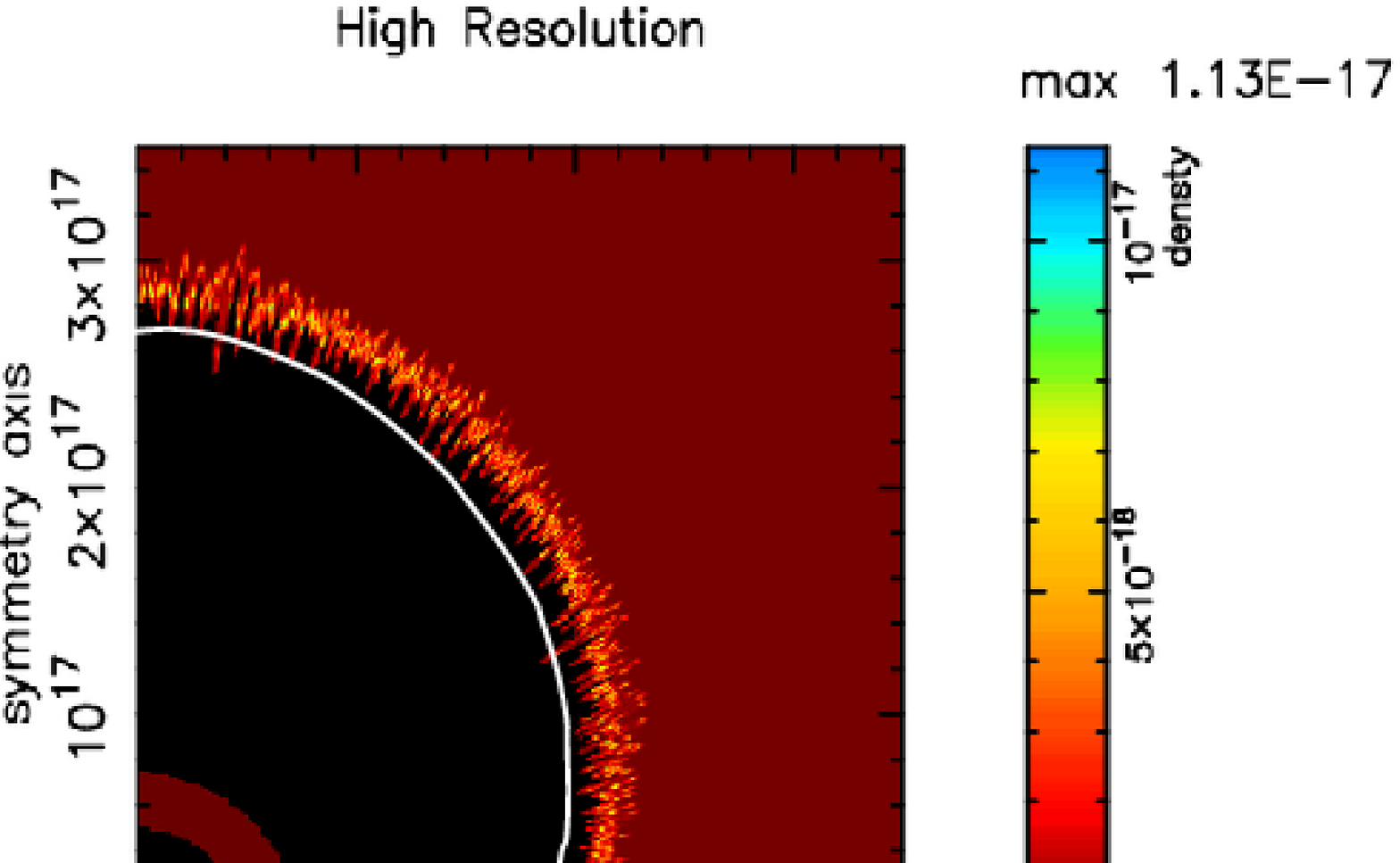}
\caption{Density profile for fully developed clumping for the
high resolution simulation. Densities are in ${\rm g\ cm^{-3}}$. The
shock is shown in white. The maximum number density achieved is $n =
3.9\times 10^6\ {\rm cm^{-3}}$. Clumping is driven by thermal
instabilities seeded by a hydrodynamic precursor. The physical
conditions assumed for the simulation are described in section
3.1. The small dark circle is the inner boundary. The low level,
declining density enhancement just outside the inner boundary is the
fast wind flowing out from the star.}
\end{figure}

\begin{figure}
\figurenum{2}
\plotone{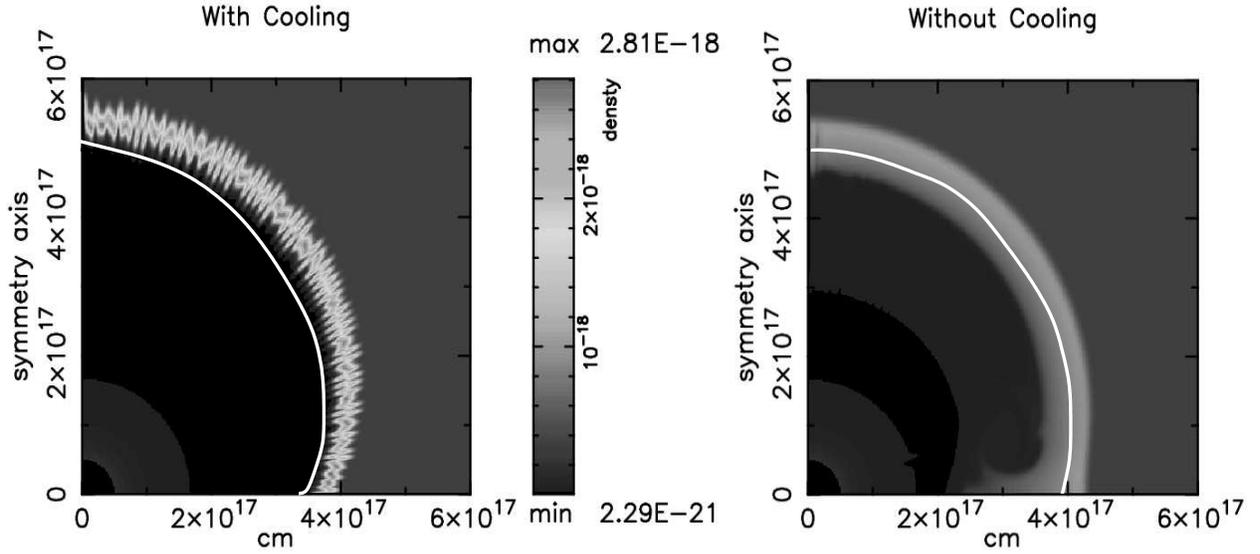}
\caption{Comparison of density profiles at similar shock radii for
cases with (left) and without (right) cooling. Without cooling the
thermal instability does not develop, and the density profile remains
smooth. (The plume at the bottom of the right hand panel is a result of
interaction of the boundary conditions with the equatorial density
enhancement. It does not appear in spherically symmetric simulations.)
The position of the shock in each case is indicated by the white
curve. The density enhancement in the case with cooling is caused by a
thermal instability driven by a hydrodynamic precursor ahead of the
shock. In the case without cooling the density enhancement is smaller
and is only due to the shock jump conditions.}
\end{figure}

\begin{figure}
\figurenum{3}
\plotone{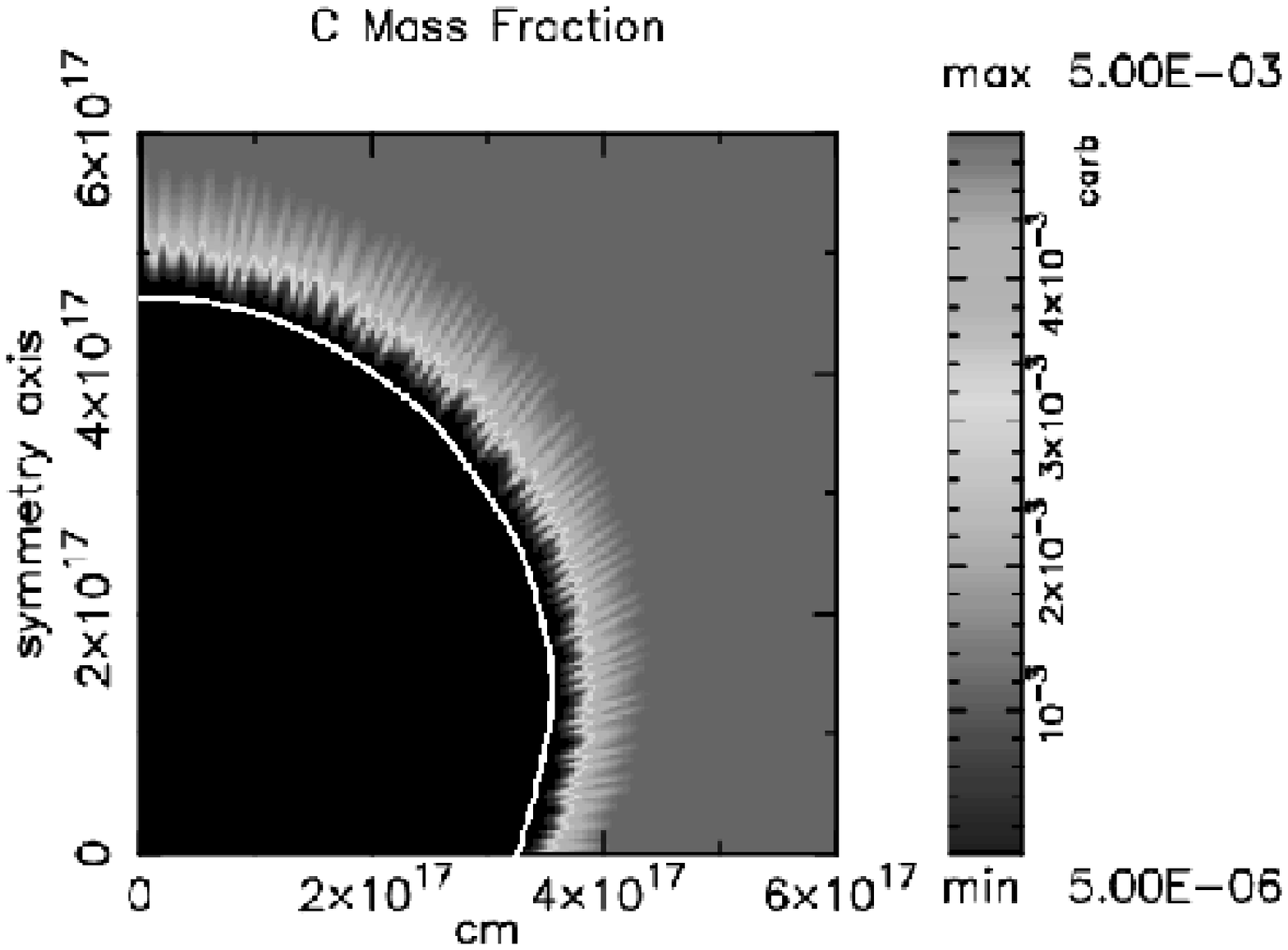}
\caption{Carbon mass fraction of the material. The C abundance of the
fast wind is set to $10^{-3}$ of the CSM to provide a Lagrangian
tracer. The position of the shock is indicated in white. The changed C
abundance ahead of the shock betrays the presence of a hydrodynamic
precursor which is driving the thermal instability. The fingering in
the C abundance closely matches that of the density profile.}
\end{figure}


\clearpage

\begin{thebibliography}


\bibitem[Cox et al.(1997)]{cox97} Cox, P., Maillard, J.-P., Huggins,
 P. J., Forveille, T., Simons, D., Guilloteau, S., Rigaut, F.,
 Bachiller, R., \& Omont, A. 1997, \aap, 321, 907

\bibitem[Cox et al.(2000)]{cox00} Cox, P., Lucas, R., Huggins, P. J.,
 Forveille, T., Bachiller, R., Guilloteau, S., Maillard, J.-P., \&
 Omont, A. 2000, \aap, 353, L25

\bibitem[Fong et al.(2003)]{fon03} Fong, D., Meixner, M., \& Shah,
R. Y. 2003, \apj, 582, L39

\bibitem[Gu\'{e}lin, Lucas, \& Neri(1997)]{gln97} Gu\'{e}lin, M.,
Lucas, R., \& Neri, R. 1997 in {\it CO: Twenty-five Years of
Millimeter Wave Spectroscopy} eds. W. B. Latter et al. (Dordrecht:
Kluwer), 359

\bibitem[Herpin et al.(2002)]{her02} Herpin, F., Goicoechea, J. R.,
Pardo, J. R., \& Cernicharo, J. 2002, \apj, 577, 961

\bibitem[Highberger et al.(2003)]{hig03} Highberger, J. L., Thomson,
K. J., Young, P. A., Arnett, D., \& Ziurys, L. M. 2003, \apj, 593, 393

\bibitem[Hollenbach \& McKee(1979)]{hm79} Hollenbach, D. \& McKee,
C. F. 1979, \apjs, 41, 555


\bibitem[Huggins et al.(2002)]{hug02} Huggins, P. J., Forveille, T.,
Bachiller, R., Cox, P., Ageorges, N., \& Walsh, J. R. 2002, \apj, 573,
L55

\bibitem[Kafatos(1973)]{kaf73} Kafatos, M. 1973, \apj, 182, 433

\bibitem[Kwok(2000)]{kwok00} Kwok, S. 2000 in {\it The Origin and
Evolution of Planetary Nebulae} (Cambridge: Cambridge University
Press)


\bibitem[Martin \& Arnett(1995)]{ma95} Martin, C. L. \& Arnett,
D. 1995, \apj, 447, 378

\bibitem[Mueller, Fryxell, \& Arnett(1991)]{mfa91} Mueller, E.,
Fryxell, B., \& Arnett, D. 1991, \aap, 251, 505

\bibitem[Press et al.(1995)]{nrc95} Press, W. H., Teukolsky, S. A.,
Vetterling, W. T., \& Flannery, B. P. 1995 {\it Numerical Recipes in
C} (New York: Cambridge University Press), 28ff.

\bibitem[O'Dell \& Handron(1996)]{oh96} O'Dell, C. R. \& Handron,
K. D. 1996, \aj, 111, 1630

\bibitem[Osterbrock(1989)]{ost} Osterbrock, D. E. 1989 {\it
Astrophysics of Gaseous Nebulae and Active Galactic Nuclei}
(Sausalito: University Science Books), 53ff.

\bibitem[Sahai et al.(1998a)]{sah98} Sahai, R., Hines, D. C., Kastner,
J. H., Weintraub, D. A., Trauger, J. T., Rieke, M. J., Thompson,
R. I., \& Schneider, G. 1998a, \apj, 492, L163

\bibitem[Sahai et al.(1998b)]{sah98b} Sahai, R. et al. 1998b, \apj,
493, 301

\bibitem[Sutherland et al.(2003)]{sut03} Sutherland, R. S., Bisset,
D. K., \& Bicknell, G. V. 2003, \apjs, 147, 187

\bibitem[Young et al.(2003)]{ykra03} Young, P. A., Knierman, K. A.,
Rigby, J. R., \& Arnett, D. 2003, \apj, in press

\bibitem[Young et al.(2001)]{ymal01} Young, P. A., 
Mamajek, E.~E., Arnett, D., \& Liebert, J. 2001, \apj, 556, 230

\bibitem[Young et al.(1992)]{you92} Young, K., Serabyn, G., Phillips,
T. G., Knapp, G. R., Guesten, R., \& Schulz, A. 1992, \apj, 385, 265


\end{thebibliography}
\end{document}